\documentclass[prb,twocolumn,showpacs,showkeys,preprintnumbers,,superscriptaddress,citeautoscript,amsmath,amssymb]{revtex4}
\usepackage{mathrsfs}
\usepackage{amssymb,amsmath}
\usepackage{graphics}
\usepackage{bm}
\usepackage{epsfig}
\usepackage{color}
\usepackage{amsfonts}
\usepackage{amscd}

\begin{document}

\title{Three-terminal thermoelectric transport through a molecule \\
 placed on an Aharonov-Bohm ring}

\author{O. Entin-Wohlman}
\email{oraentin@bgu.ac.il}

\altaffiliation{Also at Tel Aviv University.}

\affiliation{Department of Physics and the Ilse Katz Center for
Meso- and Nano-Scale Science and Technology, Ben Gurion
University, Beer Sheva 84105, Israel}

\author{A. Aharony}

\altaffiliation{Also at Tel Aviv University.}

\affiliation{Department of Physics and the Ilse Katz Center for
Meso- and Nano-Scale Science and Technology, Ben Gurion
University, Beer Sheva 84105, Israel}

\date{\today}

\begin{abstract}

The thermoelectric transport through  a ring threaded by an
Aharonov-Bohm flux, with a molecular bridge on one of its arms, is
analyzed. The charge carriers  also interact with the
vibrational excitations of that molecule. This nano-system is
connected to three terminals: two are electronic reservoirs, which
supply the charge carriers, and the third is the phonon bath
which thermalizes the molecular vibrations. Expressions for the
transport coefficients, relating all  charge and heat currents
to the temperature and chemical potential differences between the
terminals, are derived to second order in the electron-vibration
coupling. At linear response, all these coefficients obey the full
Onsager-Casimir relations. When the phonon bath is held at a
temperature different from those of the electronic reservoirs, a
heat current exchanged between the molecular vibrations and the
charge carriers  can be converted into electric and/or heat
electronic currents. The related transport coefficients, which
exist only due to the electron-vibration coupling, change sign
under the interchange between the electronic terminals and the
sign change of the magnetic flux. %At non-zero flux these effects
%can also arise for a symmetric bridge.
It is also demonstrated
that the Aharonov-Bohm flux can enhance this type of conversion.

\end{abstract}

\pacs{85.65.+h,73.63.Kv,65.80.-g}

\keywords{thermoelectric transport, charge and heat transport,
electron-vibration interaction, molecular junctions, Aharonov-Bohm
ring, Onsager-Casimir relations} \maketitle

\section{Introduction}

\label{INTRO}

Thermoelectric effects in bulk conductors usually necessitate
breaking of particle-hole symmetry. In mesoscopic structures this
asymmetry may be fairly high and can be also controlled
experimentally, leading to the possibility of realizing a
relatively large  thermopower coefficient. As a result there is currently
much interest  in  investigations of thermoelectric
phenomena   in nanoscale devices   at low temperatures.
Experimental studies have been carried out on point-contacts,
\cite{PELTIER,MOLENKAMP}  quantum dots, \cite{QDTP,PEPPER,LUDOPH}
nanotubes, \cite{COHEN,PEREZ,MCEUEN} silicon nanowires,
\cite{HOCHBAUM} and more. 
Very recently, the paramount importance of another symmetry breaking has been pointed out.
It has been  proposed that the thermal efficiency
(defined at steady state as the ratio of the output power to the  heat current)
can be significantly enhanced once {\it time-reversal
symmetry is broken and the thermopower coefficient becomes
asymmetrical} (as a function of, e.g., a magnetic field). \cite{benenti}

Early theoretical studies of
thermoelectric  transport coefficients of microstructures were
based on the Landauer approach
\cite{SIVAN,STREDA,BUTCHER,PROETTO,CBEENAKKER} and were mainly
focused on charge and heat currents between two electronic
terminals, without coupling to phonons. Later on, effects of
electron-electron processes and electronic correlations
(increasingly important at lower temperatures), as well as that of
an applied magnetic field, on the thermopower produced in large
\cite{MATVEEV} and single-level \cite{KIM} quantum dots,  and also
in  quantum wires \cite{flensberg} were considered. The signature
of attractive electronic interactions on the thermopower was
considered in Ref. ~\onlinecite{Karen}, and the dependence of the
thermoelectric response on the length of the atomic chain
connecting the leads has been recently computed within a
density-functional theory. \cite{CUEVAS}

The coupling of the charge carriers to vibrational modes of
the molecule should play a significant role in thermoelectric
transport through molecular bridges, even more so in the nonlinear
regime. \cite{FLEN} Indeed, a density-functional computation of
the nonlinear differential conductance of gold wires attributed
changes in the I-V characteristics to phonon heating,
\cite{FREDERIKSEN,NITZAN} and the thermopower coefficient was
proposed as a tool to monitor the excitation spectrum of a
molecule forming the junction between two leads. \cite{ERAN,FINCH}
It was  suggested that the Seebeck effect in such bridges can be
used for converting heat into electric energy \cite{MURPHY}, and
to determine the location of the Fermi level of the transport
electrons relative to the molecular levels, and also the sign of
the dominant charge carriers, either for a molecular conductor,
\cite{PAULSSON,TAIWAN,baranger} or for an atomic chain.
\cite{ZHENG,DVIRA} This was confirmed experimentally: the Seebeck
coefficient, as measured by STM on the benzenedithiol family
sandwiched between two gold electrodes, showed that the charge
carriers are holes passing through the HOMO, whose location with
respect to the metal Fermi level was determined from the magnitude
of the coefficient. \cite{Reddy}

Theoretically, when the coupling to the vibrational modes is
ignored, the transport coefficients have the same functional form
as in bulk conductors, with the energy-dependent transmission
coefficient and its derivative replacing  the conductivity.
\cite{CUEVAS,MAHAN} Although the corrections to the thermoelectric
transport due to the coupling to the vibrational modes is often
small, their study is  of interest because of fundamental
questions related to the symmetries of the conventional transport
coefficients, and since they give rise to additional coefficients
connecting the heat transport in-between the electrons and the
vibrational modes. In a recent article \cite{NEWWE} (referred to
below as I) we have analyzed the thermoelectric phenomena in a
molecular bridge, and studied effects induced by the coupling of
the charge carriers with  molecular vibrational modes. In
particular we have considered the situation in which the molecule
is strongly coupled to a heat bath of its own and thus may be kept
at a temperature different from those of the source and sink of
the charge carriers,  making the junction a mixed
thermal/electronic three-terminal one.   Namely, we have  assumed
that the relaxation time due to the coupling of the molecule to
its own heat bath, $\tau^{}_{\rm V}$, is short on the scale of the
coupling of the molecule to the charge carriers. The latter is
determined by the electron-vibration coupling, $\gamma$ (and the
conductance electrons density of states)  which is the small
parameter of our theory. Hence, $\hbar/ \tau^{}_{\rm V}$ may still
be very small on all other physical scales, such as
$\hbar\omega^{}_0$, where $\omega^{}_0$ is the frequency of the
vibrations, or the molecular (electronic) level width. The phonon
bath may be realized  by an electronically insulating substrate or
a piece of such material touching the junction, each held at  a
temperature $T^{}_{\rm V}$. Similar experimental three-terminal setups, consisting  of quantum dots, were discussed in  Refs. ~\onlinecite{EDWARDS},  ~\onlinecite{SL},
and ~\onlinecite{CAMB}. The effect of thermal probes  on the electronic heat conduction have been considered theoretically for tunnel normal-superconducting junctions \cite{Hekking} and for mesoscopic conductors. \cite{DAVID}

\begin{figure}[ hbtp]
\includegraphics[width=7cm]{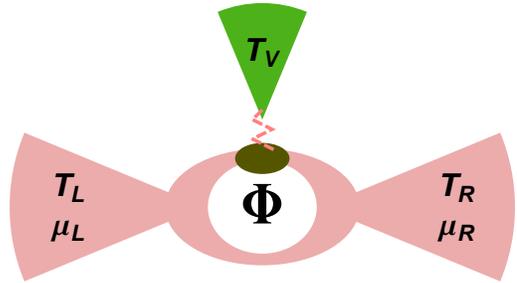}
\caption{A mixed, thermal-electronic, three-terminal system,
modeled by a localized level placed on an Aharonov-Bohm ring which
is attached to two electronic reservoirs, having different
chemical potentials and temperatures, $\mu_{\rm{L,R}}$ and
$T_{\rm{L,R}}$, respectively. An electron residing on the level
interacts with its vibrational modes. The population of these
modes can be determined  by a coupling to a phonon source kept at
temperature $T_{\rm V}$. The ring  is threaded by a magnetic flux
$\Phi$, measured in units of the flux quantum. }\label{sys}
\end{figure}

In this mixed thermal/electronic junction,  heat supplied by the
phonon baths [to which the electrons and the molecule are
(separately) coupled] can be exchanged between the transport
electrons and the vibrational modes. Even more intriguing, heat
carried from the phonon bath to the vibrations can be converted
into a charge current (or an electronic heat current)  flowing
between the electronic reservoirs. We have found in I that in
order for this conversion to take place, one needs to break the
{\em spatial} symmetry of a molecular bridge, in addition to the
broken electron-hole symmetry. Here we examine the outcome of
breaking another symmetry, that of {\it time-reversal}. To this
end, we exploit the Aharonov-Bohm effect, placing the molecular
bridge on one arm of a ring threaded by a magnetic flux, $\Phi$,
see Fig. \ref{sys}. As we show, for a non-zero flux the above
conversion can also appear in a spatially symmetric bridge, provided
that the effective couplings to the leads are energy dependent. The
specific quantities we study are the heat and the charge currents
in a small mesoscopic (or nanometric) system, depicted
schematically in Fig. \ref{sys}: a molecule (having one electronic
level) is placed on an Aharonov-Bohm  ring, which is attached to
two electronic reservoirs (held in general at different
temperatures, $T^{}_{\rm L,R}$, and at different chemical
potentials, $\mu^{}_{\rm L,R}$). The vibrational modes of the
molecule, with which the charge carriers may exchange energy,
are thermalized by a third terminal, kept at the temperature
$T^{}_{\rm V}$.

Since the calculations are rather technical, we organize the paper
as follows. After a short analysis of the entropy production in
our system (Sec. \ref{SECII}), we summarize the  main results
%of the analysis
and
in particular discuss the magnetic-flux
dependence of the transport coefficients. We then proceed in Sec.
\ref{CURS} to specify our model, and to derive the explicit
expressions for the currents. Section \ref{LRRn} details the
various transport coefficients of the three-terminal junction in
the linear-response regime. In particular, we examine the various
symmetries of the coefficients, and verify that they obey the
Onsager relations. Section \ref{DISC} is devoted to a discussion
of our results;  we consider the transport coefficients  in
certain simple cases, in particular their dependence on the
spatial and electron-hole symmetries of the junction.

\section{Entropy production and thermoelectric transport coefficients}
\label{SECII}

The consideration of  the entropy production in the
linear-response regime, carried out in I, is quite illuminating.
Using the thermodynamic identity $TdS=dE-\mu dN$, \cite{COM1} one
finds that the entropy production at the left (right) electronic
reservoir is
\begin{align}
\dot{S}^{}_{\rm L(R)}=\frac{1}{T_{\rm L(R)}}\Bigl (\dot{E}^{}_{\rm L(R)}
-\mu^{}_{\rm L(R)}\langle\dot{N}^{}_{\rm L(R)}\rangle\Bigr )\ .\label{SDL}
\end{align}
Here, $-\dot{E}_{\rm L(R)}$ is the energy current emerging from
the left (right) reservoir, while $-\langle\dot{N}_{\rm
L(R)}\rangle$ is the corresponding particle current. Adding to
Eqs. (\ref{SDL}) the entropy production of the vibrational modes,
$\dot{S}_{\rm V}=\dot{E}_{\rm V}/T_{\rm V}$, where $-\dot{E}_{\rm
V}$ is the energy current from the phonon bath into the molecule,
yields the total dissipation of the system,
\begin{align}
\dot{S}^{}_{\rm V}&+\dot{S}^{}_{\rm L}+\dot{S}^{}_{\rm R}=\frac{\dot{E}_{\rm V}}{T_{\rm V}}\nonumber\\
&+\frac{1}{T_{\rm L}}\Bigl (\dot{E}^{}_{\rm L}-\mu^{}_{\rm L}\langle\dot{N}^{}_{\rm L}\rangle\Bigr )+\frac{1}{T_{\rm R}}\Bigl (\dot{E}^{}_{\rm R}-\mu^{}_{\rm R}\langle\dot{N}^{}_{\rm R}\rangle\Bigr )\ .\label{TSD}
\end{align}
We demonstrate below  that our three-terminal junction conserves
charge, i.e., $\langle\dot{N}^{}_{\rm L}+\dot{N}^{}_{\rm
R}\rangle=0$, and also energy, i.e. $ \dot{E}^{}_{\rm
L}+\dot{E}^{}_{\rm R}+\dot{E}^{}_{\rm V}=0$. In the
linear-response regime all three temperatures (see Fig. \ref{sys})
are only slightly different from each other,
\begin{align}
T^{}_{\rm L(R)}&=T\pm\frac{\Delta T}{2}\ ,\nonumber\\
T^{}_{\rm V}&=T+\Delta T^{}_{\rm V}\ ,\label{LRT}
\end{align}
and  the chemical potentials differ by a small amount,
\begin{align}
\mu_{\rm L(R)}=\mu\pm\frac{\Delta\mu}{2}\ .\label{LRM}
\end{align}
One then finds
\begin{align}
\dot{S}^{}_{\rm V}&+\dot{S}^{}_{\rm L}+\dot{S}^{}_{\rm R}=\frac{\Delta T^{}_{\rm V}}{T^{2}}(-\dot{E}^{}_{\rm V})
+\frac{\Delta\mu /e}{T}I+\frac{\Delta T}{T^{2}}I^{}_{\rm Q}\ ,
\end{align}
where $I$ is the net charge current flowing from the left electronic reservoir to the right one,
\begin{align}
I=-\frac{e}{2}\langle\dot{N}^{}_{\rm L}-\dot{N}^{}_{\rm R}\rangle\ ,\label{I}
\end{align}
while $I_{\rm Q}$ is the net heat current carried by the electrons,
\begin{align}
I^{}_{\rm Q}=I^{}_{\rm E}-(\mu/e)I\ ,\ {\rm with}\ \ I^{}_{\rm E}=-\frac{1}{2}\Bigl (\dot{E}^{}_{\rm L}-\dot{E}^{}_{\rm R}\Bigr )\ . \label{IQ}
\end{align}
We keep the notation $-\dot{E}_{\rm V}$ for the the energy/heat
current from the phonon bath. Thus, the entropy production of our
three-terminal system is a simple example of the general
expressions for linear transport, consistent with the Onsager
theory. \cite{LL}

The calculations in the next two sections yield the relations
among the driving fields and the currents. In the linear response
limit one has
\begin{align}
\left [\begin{array}{c}I\\ I^{}_{\rm Q} \\-\dot{E}^{}_{\rm V}\end{array}\right ]
={\cal M}
\left [\begin{array}{c}\Delta\mu/e\\  \Delta T/T \\ \Delta T^{}_{\rm V}/T\end{array}\right ]\ ,\label{MAT}
\end{align}
where the matrix of the transport coefficients, ${\cal M}$, is
\begin{align}
{\cal M}=\left [\begin{array}{ccc}{\rm G}(\Phi )\ \ & \ \  {\rm K} (\Phi )\ \ & \ \ {\rm X}^{\rm V}_{}(\Phi )\\
 {\rm K}(-\Phi )\ \ &\ {\rm K}^{}_{2}(\Phi )\  & \ \ \widetilde{\rm X}^{\rm V}_{}(\Phi )\\
 {\rm X}^{\rm V}(-\Phi )\ \  & \ \ \widetilde{\rm X}^{\rm V}_{}(-\Phi )\ \ &\ \ {\rm C}_{}^{\rm V}(\Phi )\end{array}\right ]\ .
\label{M}
\end{align}
This matrix satisfies the Onsager-Casimir relations. We find below
that all three diagonal entries are even functions of the flux
$\Phi$. The off-diagonal entries of ${\cal M}$ consist each of a
term even in the flux, and another one, odd in it, obeying
altogether ${\cal M}_{ij}(\Phi ) ={\cal M}_{ji}(-\Phi )$.

The flux dependence of the transport coefficients is very
interesting. We find (see Secs. \ref{CURS} and \ref{LRRn} for
details) that there are three types of flux dependencies hiding in
those six coefficients. First, there is the one caused by
interference. The interference processes modify the self energy of
the Green functions pertaining to the Aharonov-Bohm ring, in
particular the broadening of the electronic resonance level due to
the coupling with the leads. The interference leads to terms
involving $\cos\Phi$. Secondly, there is the flux dependence which
appears in the form of $\cos 2\Phi$ (or alternatively,
$\sin^{2}\Phi$). This reflects the contributions of time-reversed
paths. These two  dependencies are even in the flux. They yield
the full  flux dependence of the diagonal entries of the matrix
${\cal M}$, Eq. (\ref{M}), and the even (in the flux) parts of the
off-diagonal elements. Finally, there is the {\em odd} dependence
in the flux, that appears as $\sin\Phi$. This dependence
characterizes the odd parts of the off-diagonal entries of ${\cal
M}$. These terms {\it necessitate} the coupling of the electrons to
the vibrational modes.

\section{The  currents}

\label{CURS}

In this section we  present our model system,  give details of the
currents' calculation,   and verify  their balance.

\subsection{The model}

In our analysis, the molecular bridge is replaced by a single
localized electronic level, representing the lowest available
orbital of the molecule; when a transport electron resides on the
level, it interacts (linearly) with Einstein vibrations. Our analysis does not include electronic interactions, but focuses on the electron-phonon ones. Thus, the  Hamiltonian of the
molecular bridge, which includes the coupling with the vibrations,
reads
\begin{align}
{\cal H}^{}_{\rm M}=\epsilon^{}_{0}c^{\dagger}_{0}c^{}_{0}+\omega^{}_{0}( b^{\dagger}b+\frac{1}{2})+\gamma (b+b^{\dagger})c^{\dagger}_{0}c^{}_{0}\ , \label{HD}
\end{align}
where $\epsilon^{}_{0}$ is the energy of the localized level,
$\omega^{}_{0}$ is the frequency of the harmonic oscillator
representing the vibrations, and $\gamma$ is its coupling to the
transport  electrons. The Hamiltonian describing the tunneling
between   the molecule and the leads is
\begin{align}
{\cal H}^{}_{\rm coup}=\sum_{k}(V^{}_{k}c^{\dagger}_{k}c^{}_{0}+{\rm Hc})+\sum_{p}(V^{}_{p}c^{\dagger}_{p}c^{}_{0}+{\rm Hc})\  \label{HCL}
\end{align}
[using $k(p)$ for the left (right) lead].
The leads' Hamiltonian is
\begin{align}
{\cal H}^{}_{\rm lead}={\cal H}^{}_{\rm L}+{\cal H}^{}_{\rm R}+{\cal H}^{}_{\rm LR}\ .
\end{align}
Here
\begin{align}
{\cal H}^{}_{\rm L(R)}=&\sum_{k(p)}\epsilon^{}_{k(p)}c^{\dagger}_{k(p)}c^{}_{k(p)}
\  \label{HL}
\end{align}
is the Hamiltonian of each of the leads
and
\begin{align}
{\cal H}^{}_{\rm LR}=\sum_{kp}V^{}_{kp}e^{i\Phi}c^{\dagger}_{k}c^{}_{\rm p}+{\rm Hc}\ \label{HDC}
\end{align}
describes the direct coupling between the two leads (pictorially
shown as the lower arm of the ring in Fig. \ref{sys}). In the
phase factor $e^{i\Phi}$, the flux $\Phi$ is measured in units of
the flux quantum, $hc/e$. Since we use units in which $\hbar =1$,
the flux quantum is $2\pi c/e$. Thus, our model Hamiltonian is
\begin{align}
{\cal H}={\cal H}_{\rm lead}+{\cal H}^{}_{\rm M}+{\cal H}^{}_{\rm coup}\ ,\label{FH}
\end{align}
where the operators $c^{\dagger}_{0}$, $c^{\dagger}_{k}$, and
$c^{\dagger}_{p}$ ($c^{}_{0}$, $c^{}_{k}$, and $c^{}_{p}$) create
(annihilate) an electron on the level, on the left lead, and on
the right lead, respectively, while $b^{\dagger}$ ($b$) creates
(annihilates) an excitation of frequency $\omega_{0}$ on the
molecule.

In the spirit of the Landauer approach,  \cite{book} the various
reservoirs connected to the leads are described by the populations
of the excitations pertaining to a specific reservoir. The
electronic reservoirs on the left and on the right of the bridge
are characterized by the electronic distributions  $f^{}_{\rm L}$
and $f^{}_{\rm R}$,
\begin{align}
f^{}_{\rm L(R)}(\omega )=\Bigl (1+\exp [\beta^{}_{\rm L(R)}(\omega
-\mu^{}_{\rm L(R)})]\Bigr )^{-1}\ , \label{FLR}
\end{align}
determined by the respective Fermi functions, with $\beta^{}_{\rm
L(R)}=1/k^{}_{\rm B}T^{}_{\rm L(R)}$. The phonon reservoir, which
determines the vibration population on the bridge, is
characterized by the Bose-Einstein distribution,
\begin{align}
N=\Bigl (\exp[\beta^{}_{\rm V}\omega^{}_{0}]-1\Bigr )^{-1}\ ,\label{NDEF}
\end{align}
with $\beta^{}_{\rm V}=1/k^{}_{\rm B}T^{}_{\rm V}$.

\subsection{Formal expressions for the currents}

The charge current leaving the left lead is given by $I^{}_{\rm
L}\equiv -e\langle\dot{N}^{}_{\rm L}\rangle$, where $N^{}_{\rm
L}=\sum_{k}c^{\dagger}_{k}c^{}_{k}$ is the number operator on the
left reservoir; $I^{}_{\rm L}$ is the sum of the currents flowing
from the left lead to the bridge and into the lower arm of the
ring, see Fig. \ref{sys}. In terms of the Keldysh Green functions,
\cite{HAUG}
 \begin{align}
I^{}_{\rm L}=-e\int\frac{d\omega}{2\pi}{\cal J}^{}_{\rm L}(\omega )\ ,\label{IL}
\end{align}
where
\begin{align}
&{\cal J}^{}_{\rm L}(\omega )=
\sum_{k} V^{}_{k}\Bigl (G^{}_{k;0}(\omega )-G^{}_{0;k}(\omega )\Bigr )^{<}_{}\nonumber\\
&+\sum_{kp} V^{}_{kp}\Bigl (e^{-i\Phi}G^{}_{k;p}(\omega
)-e^{i\Phi}G^{}_{p;k}(\omega )\Bigr )^{<}_{}\ . \label{IL1}
\end{align}
The subscripts of the Green functions indicate the corresponding
operators which form them. The electronic energy rate in the left
electronic reservoir  is  $\dot{E}^{}_{\rm L}\equiv\langle
\dot{\cal H}^{}_{\rm L}\rangle $; the energy current leaving the
left reservoir is given by Eqs. (\ref{IL}) and (\ref{IL1}), with
$e$ replaced by $\omega$ (within the integral). The electronic
currents  (charge and energy)  associated with the right lead are
$-e\langle\dot{N}^{}_{\rm R}\rangle$  and    $-\dot{E}^{}_{\rm
R}\equiv -\langle \dot{\cal H}^{}_{\rm R}\rangle $;   they   are
given by Eqs. (\ref{IL}) and (\ref{IL1}) upon interchanging $k$
with $p$, L with R,  and $\Phi$ with $-\Phi$. (The explicit
frequency dependence in some of the  following equations is
omitted for brevity.)

\subsection{Details of  the currents'  calculation}

Using the Keldysh technique, \cite{HAUG} we express the currents
[see Eqs. (\ref{IL}) and (\ref{IL1})] in terms of the Green
function of the localized level, denoted $G\equiv
G^{}_{0;0}(\omega)$. The latter is calculated up to second order
in the coupling to the vibrations, $\gamma$,
\begin{align}
G^{r}_{}&=\Bigl ([{\cal G}^{r}_{}]^{-1}_{}-\delta\epsilon^{}_{\rm V}-\Sigma^{r}_{\rm V}\Bigr )^{-1}\ ,\nonumber\\
G^{<}_{}&=G^{r}\Bigl (\Sigma^{<}_{l}+\Sigma^{<}_{\rm V}\Bigr )G^{a}_{}\ .\label{GDOTa}
\end{align}
Here, $G^{r}$ ($G^{a}$) is the retarded (advanced) Green function, and $G^{<}$ is the lesser one (below we also use  the greater Green function, $G^{>}$); ${\cal G}$ is the Green function in the absence of the coupling with the vibrations,
\begin{align}
{\cal G}^{r}_{}=\frac{1}{\omega -\epsilon^{}_{0}-\Sigma^{r}_{l}}\
,
\end{align}
where $\Sigma^r_{l}$ is the self-energy due to the coupling with
the leads, see Eq. (\ref{SIGLEAD}) below. Similarly, ${\cal
G}^<={\cal G}^r\Sigma^{<}_{l}{\cal G}^a$.
 The superscripts on the self energies $\Sigma$ have the same meaning as those indexing the Green functions.

We now discuss the two self energies which appear in Eq.
(\ref{GDOTa}). The first, which comes from the coupling to the
leads, $\Sigma^{}_{l}$, is independent of the coupling to the
vibrational mode, and is given by
\begin{align}
\Sigma^{r}_{l}=&\frac{-i}{2(1+\lambda )}\Bigl (\Gamma^{}_{\rm L}+\Gamma^{}_{\rm R}-2i\sqrt{\lambda\Gamma^{}_{\rm L}\Gamma^{}_{\rm R}}\cos\Phi \Bigr )\ ,\nonumber\\
\Sigma^{<}_{l}=&\frac{i}{(1+\lambda )^{2}}\Bigl (f^{}_{\rm
L}(\Gamma^{}_{\rm L}+\lambda\Gamma^{}_{\rm R})+
f^{}_{\rm R}(\Gamma^{}_{\rm R}+\lambda\Gamma^{}_{\rm L})\nonumber\\
&+2\sqrt{\lambda\Gamma^{}_{\rm L}\Gamma^{}_{\rm R}}(f^{}_{\rm
R}-f^{}_{\rm L})\sin\Phi \Bigr )\ .\label{SIGLEAD}
\end{align}
This self energy is determined by the resonance width
$\Gamma^{}_{\rm L}(\omega )$ [$\Gamma^{}_{\rm R}(\omega )$]
induced by the coupling with the left (right) lead, dressed by all
interference paths. In Eqs. (\ref{SIGLEAD}),  $\lambda$ is the
(dimensionless) direct coupling between the leads,
\begin{align}
\lambda (\omega )  =(2\pi )^{2}\sum_{k,p}|V^{}_{kp}|^{2}\delta
(\omega -\epsilon^{}_{k})\delta (\omega -\epsilon^{}_{p})\
.\label{lambda}
\end{align}
In our model, the transmission and reflection amplitudes of  the
lower arm of the ring alone (see  Fig. \ref{sys}),  $t^{}_{o}$ and
$r^{}_{o}$, respectively,  are chosen to be real and are given in
terms of $\lambda$,
\begin{align}
t^{2}_{o}(\omega )=\frac{4\lambda(\omega )}{[1+\lambda (\omega
)]^{2}}\ ,\ \ r^{2}_{o}(\omega )=1-t^{2}_{o}(\omega )\ .\label{TO}
\end{align}

The total width of the resonance level caused by the couplings to
the leads, $\Gamma$, is given by $-2{\rm Im} \Sigma^{r}_{l}$ [see
Eqs. (\ref{SIGLEAD}) and (\ref{lambda})],
\begin{align}
\Gamma(\omega )=\frac{\Gamma^{}_{\rm L}(\omega )+\Gamma^{}_{\rm
R}(\omega )}{1+\lambda (\omega )}\ ,
\end{align}
while the asymmetry in those couplings is characterized by
\begin{align}
\alpha ^{2}_{}(\omega )=\frac{4\Gamma^{}_{\rm L}(\omega
)\Gamma^{}_{\rm R}(\omega )}{[\Gamma^{}_{\rm L}(\omega
)+\Gamma^{}_{\rm R}(\omega )]^{2}}\ , \label{AN}
\end{align}
or alternatively
\begin{align}
\overline{\alpha}(\omega )=\frac{\Gamma^{}_{\rm L}(\omega
)-\Gamma^{}_{\rm R}(\omega )}{\Gamma^{}_{\rm L}(\omega
)+\Gamma^{}_{\rm R}(\omega )}\ , \label{ALP}
\end{align}
such that $\alpha^{2}+\overline{\alpha}^{2}=1$. None of the above details depends on the coupling
to the vibrational mode.

The second self energy in Eq. (\ref{GDOTa}), $\Sigma^{}_{\rm V}$,
comes from the coupling with the vibrations. To second order in
$\gamma$, it is given by
\begin{align}
\Sigma^{<}_{\rm V}(\omega )&=\gamma^{2}\Bigl (N{\cal G}^{<}_{}(\omega -\omega^{}_{0})+(1+N){\cal G}^{<}_{}(\omega +\omega^{}_{0})
\Bigr )\nonumber\\
\Sigma^{>}_{\rm V}(\omega )&=\gamma^{2}\Bigl (N{\cal G}^{>}_{}(\omega +\omega^{}_{0})+(1+N){\cal G}^{>}_{}(\omega -\omega^{}_{0})
\Bigr )\ ,\label{SIGPa}
\end{align}
where $N$ denotes the population of the vibrational modes [see Eq.
(\ref{NDEF})];   $\delta\epsilon^{}_{\rm V}$ is the polaron shift,
\begin{align}
\delta\epsilon^{}_{\rm V} =2i\frac{\gamma^{2}}{\omega^{}_{0}}\int\frac{d\omega}{2\pi}{\cal G}^{<}_{}(\omega)\ .
\end{align}
Note that the Green functions here contain $\Sigma^<_l$ and
$\Sigma^>_l$, and therefore they also depend on the flux $\Phi$.

In terms of these parameters, the integrand determining the
electronic currents running from left to right  [see Eqs.
(\ref{IL1}) and (\ref{GDOTa}),  and Fig. \ref{sys}] is
\begin{widetext}
\begin{align}
{\cal J}^{}_{\rm L}=&(f^{}_{\rm R}-f^{}_{\rm L})\Bigl (t^{2}_{o}[1-\Gamma{\rm Im}G^{a}_{}]+t^{}_{o}r^{}_{o}\Gamma\alpha\cos\Phi{\rm Re}G^{a}_{}\Bigr )\nonumber\\
&-i\frac{\Gamma}{2}\Bigl ((1+r^{}_{o}\overline{\alpha})[G_{}^{<}-f^{}_{\rm L}(G^{a}_{}-G^{r}_{})]+\alpha t^{}_{o}\sin\Phi[G_{}^{<}-f^{}_{\rm R}(G^{a}_{}-G^{r}_{})]\Bigr )\ .\label{LCURI}
\end{align}
\end{widetext}
The corresponding integrand determining the currents flowing from right to left, ${\cal J}_{\rm R}$,  is obtained from Eq. (\ref{LCURI}) upon interchanging L with R and $\Phi$ with $-\Phi$, a transformation which leaves $G$ invariant.

Let us first consider the {\em sum} of the electronic currents,
flowing from the left towards the ring and from the right towards
the ring.  Using Eqs. (\ref{SIGLEAD}), we find
\begin{align}
{\cal J}^{}_{\rm L}+{\cal J}^{}_{\rm R}&=(\Sigma^{r}_{l}-\Sigma^{a}_{l})G^{<}_{}+\Sigma^{<}_{l}(G^{a}_{}-G^{r}_{})\nonumber\\
&=|G^{a}_{}|^{2}\Bigl (\Sigma^{>}_{l}\Sigma^{<}_{\rm
V}-\Sigma^{<}_{l}\Sigma^{>}_{\rm V}\Bigr )\ .
\end{align}
Therefore,
up to second order in the coupling with the vibrations, the sum of the currents is given by
\begin{align}
&\int\frac{ \omega^{s}_{}d\omega}{2\pi}\Bigl ({\cal J}^{}_{\rm
L}+{\cal J}^{}_{\rm R}\Bigr ) =
\int\frac{\omega^{s}_{}d\omega}{2\pi}|{\cal G}_{}^{a}|^{2}_{}\Bigl
(\Sigma^{>}_{l}\Sigma_{\rm V}^{<}-\Sigma^{<}_{l}\Sigma^{>}_{\rm
V}\Bigr )\ , \label{SUM}
\end{align}
where $s=0$ for the sum of the two electronic electric currents, and $s=1$ for the sum of the two  electronic energy currents.
Inserting  here Eqs. (\ref{SIGPa}) for the self energy due to the coupling with the vibrations,
one obtains that Eq. (\ref{SUM})  vanishes for $s=0$, ensuring the electric current conservation,
\begin{align}
\langle \dot{N}^{}_{\rm L}\rangle +\langle\dot{N}^{}_{\rm R}\rangle =0\ .
\end{align}
On the other hand, for $s=1$ one finds
\begin{align}
&\int\frac{\omega d\omega}{2\pi} \Bigl ({\cal J}^{}_{\rm L}(\omega )+{\cal J}^{}_{\rm R}(\omega )\Bigr )=
\gamma^{2}\omega^{}_{0}\int\frac{d\omega}{2\pi}|{\cal G}^{a}_{}(\omega ^{}_{-}){\cal G}^{a}_{}(\omega ^{}_{+})|^{2}\nonumber\\
&\times\Bigl (N\Sigma^{>}_{l}(\omega^{}_{+})\Sigma^{<}_{l}(\omega ^{}_{-})-(1+N)\Sigma^{>}_{l}(\omega^{}_{-})\Sigma^{<}_{l}(\omega ^{}_{+})\Bigr )\ ,\label{PCUR}
\end{align}
where we have introduced the abbreviations
\begin{align}
\omega^{}_{\pm}=\omega\pm\frac{\omega^{}_{0}}{2}\ .\label{OMPM}
\end{align}
The energy current carried by the vibrations, $\omega_{0}d\langle b^{\dagger}b\rangle/dt$ [see Eq. (\ref{HD})], may be calculated from the Keldysh phonon Green functions,  \cite{HAUG} and it is straightforward to show  that it is given by the result
(\ref{PCUR}),
ensuring the total energy conservation of our model,
\begin{align}
-\dot{E}^{}_{\rm V}=\dot{E}^{}_{\rm L}+\dot{E}^{}_{\rm R}\ .\label{EP}
\end{align}

\subsection{Explicit expressions for the currents}

Returning to Eq. (\ref{LCURI}) and inserting there Eqs.
(\ref{GDOTa}), we find that, like the self energies, ${\cal
J}_{\rm L}$ can be separated into  two contributions, one from the
coupling to the leads and the other from the coupling with the
vibrations,
\begin{align}
{\cal J}^{}_{\rm L}(\omega )=&[f^{}_{\rm R}(\omega )-f^{}_{\rm
L}(\omega )]{\cal J}^{l}_{}(\omega ) +{\cal J}^{\rm V}_{\rm
L}(\omega )\ ,\label{CURLL}
\end{align}
where the `bare' transition probability between the leads is (cf.
Ref. ~\onlinecite{HOFSTETTER})
\begin{align}
{\cal J}^{l}_{}(\omega )&=t^{2}_{o}\Bigl (1-\Gamma{\rm Im}G^{a}_{}+\frac{\Gamma^{2}}{4}[1-\alpha^{2}_{}\cos^{2}\Phi ]|G^{a}_{}|^{2}\Bigr )\nonumber\\
&+t^{}_{o}r^{}_{o}\Gamma\alpha\cos\Phi{\rm Re}G^{a}_{}
+\frac{\Gamma^{2}\alpha^{2}}{4}|G^{a}_{}|^{2}\ .\label{ELTR}
\end{align}
The first term here, $t^{2}_{o}$,  yields the conductance in the
absence of the ring arm carrying the bridge (see Fig. \ref{sys}),
while the last term yields the conductance of that arm alone. All
other three terms in Eq. (\ref{ELTR}) result from various
interference processes which do not involve the coupling to the
vibrations [apart from their renormalization of the bridge Green
function, see Eq. (\ref{GDOTa})]. This `bare' transition probability  is an {\em
even} function of the flux $\Phi$.

The contribution of the coupling to the vibrations, i.e. the last
term of Eq. (\ref{CURLL}), is given to order $\gamma^2$ by
\begin{align}
{\cal J}^{\rm V}_{\rm L}(\omega )
&=-i\frac{\Gamma }{2}|{\cal G}^{a}_{}|^{2}\Bigl (a^{}_{+}[\Sigma^{<}_{\rm V}(1-f^{}_{\rm L})+\Sigma^{>}_{\rm V}f^{}_{\rm L}]\nonumber\\
&+b\sin\Phi[\Sigma^{<}_{\rm V}(1-f^{}_{\rm R})+\Sigma^{>}_{\rm
V}f^{}_{\rm R}]\Bigr )\ , \label{LCURIN}
\end{align}
where
\begin{align}
a^{}_{\pm}(\omega )&=1\pm\overline{\alpha}(\omega )r^{}_{o}(\omega )\ ,\nonumber\\
b(\omega )&=\alpha (\omega )t^{}_{o}(\omega )\ ,\label{AB}
\end{align}
see Eqs. (\ref{TO}), (\ref{AN}), and  (\ref{ALP}).  (The function $a_{-}$ is obtained from the function $a_{+}$ upon interchanging L with R.)

We are now in position to present the detailed expressions for the currents,
by inserting Eq. (\ref{CURLL}) [and the analogous one, obtained upon interchanging
there L with R and $\Phi$ with $-\Phi$] into Eqs. (\ref{I}), (\ref{IQ}), and (\ref{EP}). The electric current is
\begin{align}
I&=-e\int\frac{d\omega}{2\pi}[f^{}_{\rm R}(\omega)-f^{}_{\rm
L}(\omega )]{\cal J}^{l}_{}(\omega )
\nonumber\\
&-e\int\frac{d\omega}{4\pi}c(\omega,\Phi ){\cal J}^{\rm
V}_{0}(\omega )\ ,\label{IFINn}
\end{align}
while the energy current carried by the electrons is
\begin{align}
&I^{}_{\rm E}=-\int\frac{\omega d\omega}{2\pi}
[f^{}_{\rm R}(\omega)-f^{}_{\rm L}(\omega )]{\cal J}^{l}_{}(\omega )\nonumber\\
& -\int\frac{ d\omega}{4\pi}c(\omega,\Phi ){\cal J}^{\rm V}_{1}(\omega
) \ .\label{IEFINn}
\end{align}
The function
\begin{align}
c(\omega , \Phi)=\gamma^{2}\frac{\Gamma (\omega^{}_{+})\Gamma (\omega^{}_{-})}{4}|{\cal G}^{a}_{}(\omega^{}_{+}){\cal G}^{a}_{}(\omega^{}_{-})|^{2}\label{cf}
\end{align}
introduced in Eqs. (\ref{IFINn}) and (\ref{IEFINn}) has a simple
meaning. Keeping in mind that $(\Gamma /2)|{\cal G}^{a}|^{2}$ sets
the scale for the local density of states on the bridge (to
leading order), it is seen that $c(\omega, \Phi)$ is  the product
of the local  densities of states at the two shifted frequencies
$\omega_{\pm}$ [see Eq. (\ref{OMPM})], multiplied by the coupling
to the vibrations, $\gamma^{2}$. Note that the broadening of the
energy level on the dot resulting from the coupling to the leads
depends on the flux $\Phi$, see Eqs. (\ref{SIGLEAD}), leading in
turn to the flux dependence of   ${\cal G}$ and the function $c$,
Eq. (\ref{cf}). This is the $\cos\Phi$ dependence resulting from
the usual interference processes, discussed at the end of Sec.
\ref{SECII}.

The other function introduced above is ${\cal J}^{\rm V}_{s}$,
where $s=0$ in Eq. (\ref{IFINn}), and $s=1$ in Eq. (\ref{IEFINn}),
\begin{widetext}
\begin{align}
{\cal J}^{\rm V}_{s}(\omega )
&=F^{}_{\rm LL}(\omega )\Bigl (\omega^{s}_{+}g^{}_{++}(\omega^{}_{+})g^{}_{+-}(\omega ^{}_{-})-\omega^{s}_{-}g^{}_{++}(\omega^{}_{-})g^{}_{+-}(\omega ^{}_{+})\Bigr )\nonumber\\
&+F^{}_{\rm RR}(\omega )\Bigl (-\omega^{s}_{+}g^{}_{--}(\omega^{}_{+})g^{}_{-+}(\omega ^{}_{-})+\omega^{s}_{-}g^{}_{--}(\omega^{}_{-})g^{}_{-+}(\omega ^{}_{+})\Bigr )\nonumber\\
&+F^{}_{\rm LR}(\omega )\Bigl (\omega^{s}_{+}g^{}_{++}(\omega^{}_{+})g^{}_{-+}(\omega ^{}_{-})+\omega^{s}_{-}g^{}_{--}(\omega^{}_{-})g^{}_{+-}(\omega ^{}_{+})\Bigr )\nonumber\\
&+F^{}_{\rm RL}(\omega )\Bigl (-\omega^{s}_{+}g^{}_{--}(\omega^{}_{+})g^{}_{+-}(\omega ^{}_{-})-\omega^{s}_{-}g^{}_{++}(\omega^{}_{-})g^{}_{-+}(\omega ^{}_{+})\Bigr )\ .
\end{align}
\end{widetext}
The various excitation populations characterizing the reservoirs
are incorporated into the functions $F^{}_{nn'}$,
\begin{align}
F^{}_{\rm nn'}(\omega )&=N[1-f^{}_{\rm n}(\omega ^{}_{+})]f^{}_{\rm n'}(\omega^{}_{-})\nonumber\\
&-(1+N)[1-f^{}_{\rm n'}(\omega^{}_{-})]f^{}_{\rm n}(\omega^{}_{+})\ , \label{FFn}
\end{align}
[see Eqs. (\ref{FLR}) and (\ref{NDEF})]. In addition, the four functions $g_{\pm\pm}$ are given by
\begin{align}
g^{}_{\pm \pm}(\omega )=a^{}_{\pm}(\omega )\pm b(\omega)\sin\Phi\
\end{align}
for the equal-sign functions $g$,  and
\begin{align}
g^{}_{\pm\mp}(\omega )=a^{}_{\pm}(\omega )\mp b(\omega) \sin\Phi\
\end{align}
for the unequal ones.

Finally, the energy current carried by the vibrations is given by
\begin{align}
-\dot{E}^{}_{\rm V}=&\omega^{}_{0}\int\frac{d\omega}{4\pi}\Bigl (F^{}_{\rm LL}(\omega )g^{}_{+ -}(\omega^{}_{+})g^{}_{+ -}(\omega^{}_{-})\nonumber\\
&+F^{}_{\rm RR}(\omega )g^{}_{-+}(\omega^{}_{+})g^{}_{-+}(\omega^{}_{-})\nonumber\\
&+F^{}_{\rm LR}(\omega )g^{}_{+-}(\omega^{}_{+})g^{}_{-+}(\omega^{}_{-})\nonumber\\
&+F^{}_{\rm RL}(\omega )g^{}_{-+}(\omega^{}_{+})g^{}_{+-}(\omega^{}_{-})\Bigr )\ .
\end{align}
These lengthy expressions are significantly simplified when one
assumes a symmetrically-coupled bridge, or a perfect bare
transmission of the lower arm of the ring (see Fig. \ref{sys}). In
both cases,  Eqs. (\ref{AB}) yield  $a_{\pm}=1$.

\section{The linear-response regime}

\label{LRRn}

In the linear-response regime  one expands all currents to first
order in the driving forces, which in our case are $\Delta\mu$,
$\Delta T$, and $\Delta T_{\rm V}$ (see discussion in Sec.
\ref{SECII}). This amounts to expanding the Fermi distributions,
Eqs. (\ref{FLR}),  around the common temperature $T$ and the
common chemical potential $\mu$ of the device, i.e., around
$f(\omega )=[\exp(\beta (\omega -\mu ))+1]^{-1}$, and the
vibrational mode population, Eq. (\ref{NDEF}), around the Bose
function $N^{}_{\rm T}=[\exp(\beta \omega ^{}_{0} )-1]^{-1}$,
where $\beta=1/k^{}_BT$.

The electric current in the linear-response regime takes the form
\begin{align}
I={\rm G}(\Phi )\frac{\Delta\mu}{e}+{\rm K}(\Phi )\frac{\Delta T}{T}+{\rm X}^{\rm V}_{}(\Phi )\frac{\Delta T^{}_{\rm V}}{T}\ .  \label{ILINn}
\end{align}
As is clear from the  discussion in Sec. \ref{CURS}, there are two
contributions to each of the first two terms of the  electric
current, one resulting from the coupling to the leads [the first
term on the right-hand side of Eq. (\ref{IFINn})] and the other
coming from the coupling to the vibrations. The last term in Eq.
(\ref{ILINn}) arises from the latter coupling alone. Accordingly,
\begin{align}
{\rm G}&={\rm G}^{l}_{}+{\rm G}^{\rm V}_{}\ ,\nonumber\\
{\rm K}&={\rm K}^{l}_{}+{\rm K}^{\rm V}_{}\ ,
\end{align}
with
\begin{align}
{\rm G}^{l}_{}(\Phi )=e^{2}\int\frac{d\omega}{2\pi}{\cal
T}^{l}_{}(\omega, \Phi )\ ,
\end{align}
where
\begin{align}
{\cal T}^{l}_{}(\omega, \Phi )=\beta f(\omega )[1-f(\omega )]{\cal
J}^{l}_{}(\omega, \Phi )\
\end{align}
is essentially the transition probability, Eq. (\ref{ELTR}),  at the Fermi energy (at low temperatures). Similarly the transport coefficient related to the thermopower is
\begin{align}
{\rm K}^{l}_{}(\Phi )=e\int\frac{d\omega }{2\pi}(\omega -\mu
){\cal T}^{l}_{}(\omega, \Phi )\ .
\end{align}
These expressions for the transport coefficients are  the standard
ones, see for example Ref. ~\onlinecite{SIVAN}; both coefficients,
${\rm G}^{l}$ and ${\rm K}^{l}$, are even in the magnetic flux.

Next we examine the contribution of the `vibronic' transitions to
the transport coefficients. Introducing the function
\begin{align}
{\cal T}^{\rm V}_{}(\omega, \Phi )=\beta N^{}_{\rm
T}f(\omega^{}_{-} )[1-f(\omega^{}_{+} )]c(\omega, \Phi )\
,\label{TINEL}
\end{align}
we find
\begin{align}
{\rm G}^{\rm V}_{}(\Phi )=e^{2}\int\frac{d\omega}{\pi}{\cal
T}^{\rm V}_{}(\omega ,\Phi )m^{}_{0}(\omega ,\Phi )\ ,
\end{align}
where
\begin{align}
m^{}_{0}(\omega ,\Phi )&=1-\overline{\alpha}(\omega^{}_{-})\overline{\alpha}(\omega^{}_{+})r^{}_{o}(\omega ^{}_{-})r^{}_{o}(\omega^{}_{+})\nonumber\\
&+\alpha(\omega^{}_{-})\alpha(\omega^{}_{+})t^{}_{o}(\omega ^{}_{-})t^{}_{o}(\omega^{}_{+})\sin^{2}\Phi\ .
\end{align}
[We remind the reader that $\alpha$ and $\overline{\alpha}$
characterize the asymmetry of the bridge, see Eqs. (\ref{ALP}),
while $t^{}_{o}$ and $r^{}_{o}$ are the transmission and
reflection amplitudes of the lower arm of the ring, see Eqs.
(\ref{TO}). The frequencies $\omega_{\pm}$ are given in Eq. (\ref{OMPM}).] Notice  that the full conductance G$^{l}$+G$^{\rm V}$
is even in the flux. Thus it is seen that the electron-vibration
interaction introduces only a modest modification in the
conductance, but does not lead to major effects.

This is not the case with all other transport coefficients of the
ring. The thermopower coefficient ${\rm K}^{\rm V}$ is given by
\begin{align}
{\rm K}^{\rm V}_{}(\Phi )&=e\int\frac{d\omega}{\pi}(\omega -\mu ){\cal T}^{\rm V}_{}(\omega ,\Phi )m^{}_{0}(\omega ,\Phi )\nonumber\\
&+e\frac{\omega^{}_{0}}{2}\sin\Phi \int\frac{d\omega}{\pi}{\cal
T}^{\rm V}_{}(\omega ,\Phi )m^{}_{1}(\omega  )\ ,
\end{align}
where
\begin{align}
m^{}_{1}(\omega )&=\alpha(\omega^{}_{+})t^{}_{o}(\omega ^{}_{+})\overline{\alpha}(\omega^{}_{-})r^{}_{o}(\omega ^{}_{-})\nonumber\\
&-
\alpha(\omega^{}_{-})t^{}_{o}(\omega ^{}_{-})\overline{\alpha}(\omega^{}_{+})r^{}_{o}(\omega ^{}_{+})\ .
\end{align}
While the first term of  ${\rm K}^{\rm V}$ is  even in the flux,
the second term is an odd function of $\Phi$. In order not to
vanish, this odd component  necessitates that the bridge will not
be coupled symmetrically to the leads, and that the transmission
of the lower arm of the ring will not be perfect (to ensure
interference).  In addition, ${\rm K}^{\rm V}$ is invariant under
the transformation L$\leftrightarrow$R [which changes the sign of
$\overline{\alpha}$, see Eq. (\ref{ALP})]  and
$\Phi\leftrightarrow -\Phi$.

Finally, the coefficient ${\rm X}^{\rm V}$  [see Eq. (\ref{ILINn})] is
\begin{align}
{\rm X}^{\rm V}_{}(\Phi )=e\omega^{}_{0}\int\frac{d\omega }{\pi}
{\cal T}^{\rm V}_{}(\omega ,\Phi )m^{}_{2}(\omega  ,\Phi)\
,\label{XP}
\end{align}
with
\begin{align}
m^{}_{2}(\omega ,\Phi)&=\overline{\alpha}(\omega^{}_{-})r^{}_{o}(\omega ^{}_{-})+\alpha(\omega^{}_{-})t^{}_{o}(\omega ^{}_{-})\sin\Phi\nonumber\\
&-\overline{\alpha}(\omega^{}_{+})r^{}_{o}(\omega ^{}_{+})-\alpha(\omega^{}_{+})t^{}_{o}(\omega ^{}_{+})\sin\Phi
\ .\label{m2}
\end{align}
Note that under the transformation L$\leftrightarrow$R and
$\Phi\leftrightarrow -\Phi$, ${\rm X}^{\rm V}$ changes its sign.
Also note that, again, the $\sin\Phi$ dependence of this transport
coefficient necessitates the coupling to the vibrational modes.

In an analogous fashion we derive the heat current carried by the electrons, and find
\begin{align}
I^{}_{\rm Q}={\rm K}(-\Phi )\frac{\Delta\mu}{e}+{\rm K}^{}_{2}(\Phi )\frac{\Delta T}{T}+\widetilde{\rm X}^{\rm V}_{}(\Phi )\frac{\Delta T^{}_{\rm V}}{T}\ . \label{IELINn}
\end{align}
Here, the transport coefficient related to the thermal
conductivity, ${\rm K}_{2}$, has again two contributions,
\begin{align}
{\rm K}^{}_{2}&={\rm K}^{l}_{2}+{\rm K}^{\rm V}_{2}\ ,
\end{align}
where
\begin{align}
{\rm K}^{l}_{2}(\Phi )=\int\frac{d\omega }{2\pi}(\omega -\mu
)^{2}{\cal T}^{l}_{}(\omega , \Phi )\
\end{align}
as expected, while the `vibronic' term is
\begin{align}
{\rm K}^{\rm V}_{2}(\Phi )&=\int\frac{d\omega}{\pi}(\omega -\mu)^{2}{\cal T}^{\rm V}_{}(\omega ,\Phi )m^{}_{0}(\omega ,\Phi )\nonumber\\
&+\frac{\omega^{2}_{0}}{4}\int\frac{d\omega}{\pi}{\cal T}^{\rm
V}_{}(\omega ,\Phi )\overline{m}^{}_{0}(\omega ,\Phi )\ ,
\end{align}
with
\begin{align}
\overline{m}^{}_{0}(\omega ,\Phi )&=1+\overline{\alpha}(\omega^{}_{-})\overline{\alpha}(\omega^{}_{+})r^{}_{o}(\omega ^{}_{-})r^{}_{o}(\omega^{}_{+})\nonumber\\
&-\alpha(\omega^{}_{-})\alpha(\omega^{}_{+})t^{}_{o}(\omega ^{}_{-})t^{}_{o}(\omega^{}_{+})\sin^{2}\Phi\ .
\end{align}
One notes that ${\rm K}_{2}$ is even in the flux. The last transport coefficient in Eq. (\ref{IELINn}) is
\begin{align}
\widetilde{\rm X}^{\rm V}_{}(\Phi )&=\omega^{}_{0}\int\frac{d\omega}{\pi}(\omega -\mu )^{}{\cal  T}^{\rm V}_{}(\omega ,\Phi )m^{}_{2}(\omega  ,\Phi)\nonumber\\
&-\frac{\omega^{2}_{0}}{2}\int\frac{d\omega}{\pi}{\cal  T}^{\rm
V}_{}(\omega ,\Phi )\overline{m}^{}_{2}(\omega  ,\Phi)\
,\label{XPT}
\end{align}
with
\begin{align}
\overline{m}^{}_{2}(\omega ,\Phi)&=\overline{\alpha}(\omega^{}_{-})r^{}_{o}(\omega ^{}_{-})+\alpha(\omega^{}_{-})t^{}_{o}(\omega ^{}_{-})\sin\Phi\nonumber\\
&+\overline{\alpha}(\omega^{}_{+})r^{}_{o}(\omega ^{}_{+})+\alpha(\omega^{}_{+})t^{}_{o}(\omega ^{}_{+})\sin\Phi
\ .\label{m2k}
\end{align}
This coefficient changes its sign under the transformation
L$\leftrightarrow$R and $\Phi\leftrightarrow -\Phi$, as expected
from the Onsager relations.

The energy current from the thermal bath takes the form
\begin{align}
-\dot{E}^{}_{\rm V}={\rm X}^{\rm V}_{}(-\Phi )\frac{\Delta\mu}{e}+\widetilde{\rm X}^{\rm V}_{}(-\Phi )\frac{\Delta T}{T}+{\rm C}^{\rm V}_{}\frac{\Delta T^{}_{\rm V}}{T}\ ,
\end{align}
where the third coefficient here is
\begin{align}
{\rm C}^{\rm V}_{}=2\omega^{2}_{0}\int\frac{d\omega}{\pi}{\cal
T}^{\rm V}_{}(\omega ,\Phi )\ ,
\end{align}
and is an even function of the flux.
Collecting all results, we arrive at the matrix form for the relations among the currents and the driving forces, Eqs. (\ref{MAT}) and (\ref{M}).

\section{Examples and discussion}

\label{DISC}

As is clear from the results in Sec. \ref{LRRn} (and also from
Ref. ~\onlinecite{NEWWE}, which treated the bridge without the
lower branch), the interesting effect induced exclusively  by the
coupling of the electrons to the vibrational modes is the
possibility to create an electric current, or an electronic heat
current,  by applying a temperature difference $\Delta T^{}_{\rm
V}$ on the phonon bath thermalizing this mode. These new
thermoelectric phenomena are specified by the two coefficients
 ${\rm X}^{\rm V}$ and  $\widetilde{\rm X}^{\rm V}$,
Eqs. (\ref{XP})  and  (\ref{XPT}), respectively.  All other
transport coefficients related to the electronic currents are
mainly due to the transport of the electrons between the
electronic terminals, with slight modifications from the (small)
coupling to the vibrations. We therefore confine the main
discussion in this section to the coefficients
 ${\rm X}^{\rm V}$ and  $\widetilde{\rm X}^{\rm V}$. To make a closer
 connection with a possible experiment, we introduce the (dimensionless)  coefficients
\begin{align}
{\rm S}^{\rm V}_{}=e\beta\frac{ {\rm X}^{\rm V}_{}}{\rm G}\ ,
\end{align}
and
\begin{align}
\widetilde{S}^{\rm V}_{}=\frac{\widetilde {\rm X}^{\rm V}_{}}{\rm K_{2}^{}}\ .
\end{align}
The first gives the potential drop across the molecular bridge
created by $\Delta T^{}_{\rm V}$ when the temperature drop there,
$\Delta T$, vanishes, and the second yields the temperature
difference created by $\Delta T^{}_{\rm V}$ when $\Delta\mu=0$
[or the inverse processes, see Eqs. (\ref{MAT}) and (\ref{M})].
For both the conductance, G, and the thermal conductance, K$_{2}$,
we use below their leading terms, resulting from the coupling to
the leads  alone (the numerators result from the coupling to the vibrations, and hence are already of order $\gamma^2$).

As  mentioned above, the transport coefficients of our
three-terminal junction obey the Onsager-Casimir relations. They
do it  though in a somewhat unique way:  the ``off-diagonal"
elements are related to one another by the reversal of the
magnetic field. However, they are not a purely odd function of it.  A
special situation arises when the molecule is connected {\em
symmetrically} to the two leads. In that case, the anisotropy
parameter $\overline{\alpha}$ vanishes, while $\alpha =1$ [see
Eqs. (\ref{AN}) and (\ref{ALP})]. Then, the two coefficients,
${\rm X}^{\rm V}$ and  $\widetilde{\rm X}^{\rm V}$,  are odd
functions of the flux, resembling the thermal Hall effect
discussed recently in connection with quantum magnets,  \cite{LEE}
\begin{align}
{\rm X}^{\rm V}_{}(\Phi )=e\omega^{}_{0}\sin \Phi
\int\frac{d\omega }{\pi} {\cal T}^{\rm V}_{}(\omega ,\Phi )\Bigl
(t^{}_{o}(\omega^{}_{-})-t^{}_{o}(\omega^{}_{+})\Bigr )\
,\label{XP1}
\end{align}
and
\begin{align}
&\widetilde{\rm X}^{\rm V}_{}(\Phi )=\omega^{}_{0}\sin \Phi \int\frac{d\omega}{\pi}{\cal  T}^{\rm V}_{}(\omega ,\Phi )\nonumber\\
&\times\Bigl (t^{}_{o}(\omega ^{}_{-})(\omega^{}_{-}-\mu )-t^{}_{o}(\omega ^{}_{+})(\omega^{}_{+}-\mu)\Bigr )\ .
\label{XPT1}
\end{align}
In other words, the thermoelectric processes described by ${\rm
X}^{\rm V}_{}(\Phi )$ and $\widetilde{\rm X}^{\rm V}_{}(\Phi )$
require a certain symmetry-breaking. In the absence of the
magnetic field, that is supplied by the spatial asymmetry of the
junction; in the presence of a flux, those processes appear also
for a  junction symmetrically coupled to the leads, provided that the couplings
to the leads depend on the energy.

\begin{figure}[ hbtp]
\includegraphics[width=6.cm]{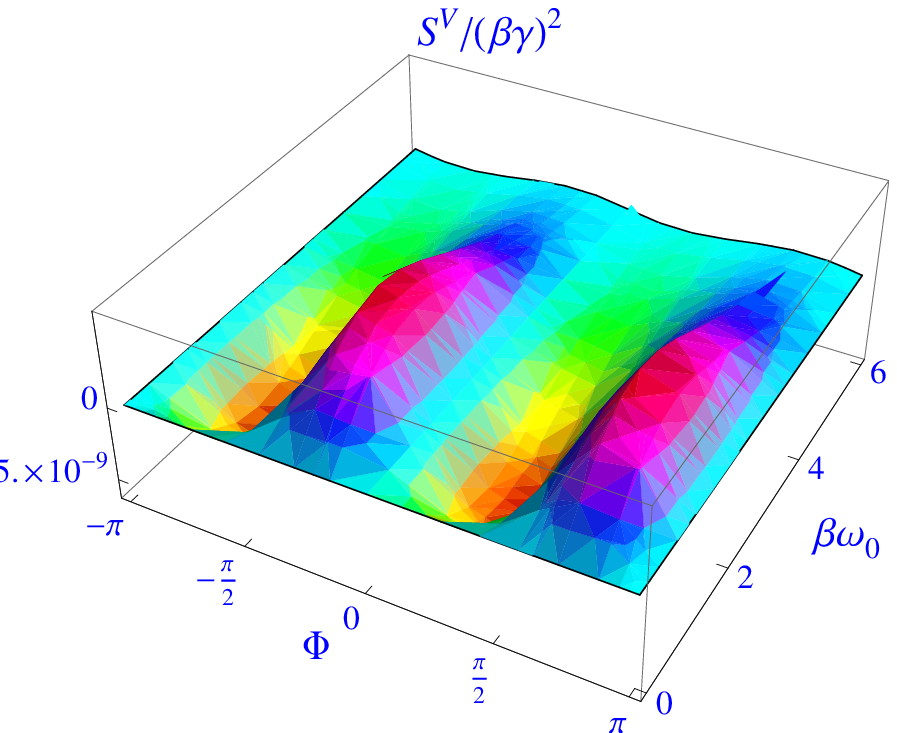}
\includegraphics[width=6cm]{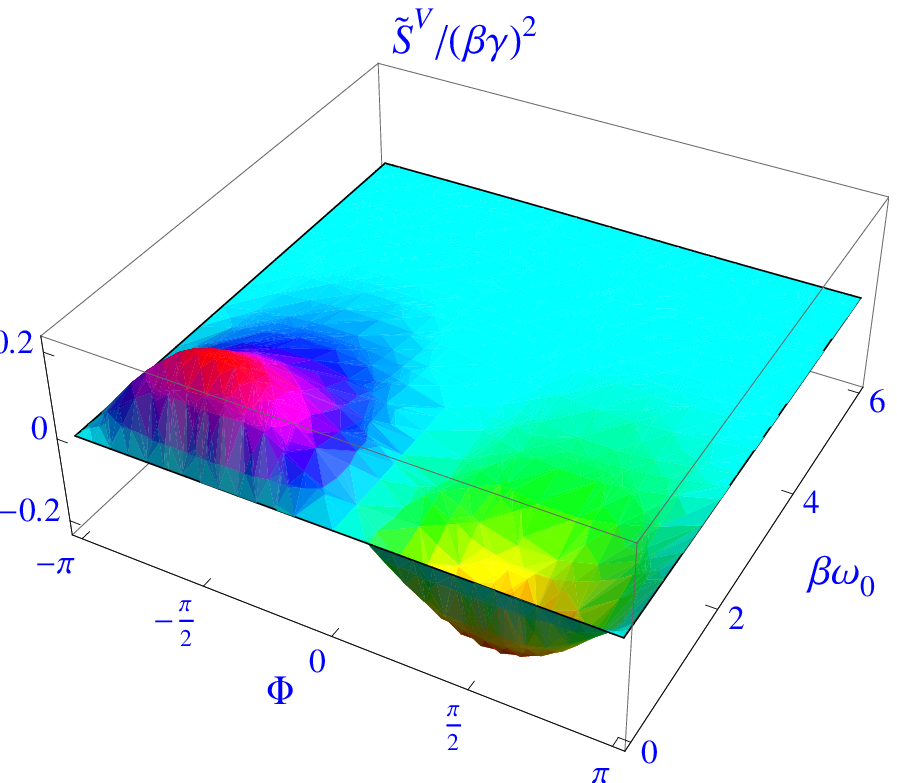}\caption{(color online.) The
transport coefficients ${\rm S}^{\rm V}$ and $\widetilde{\rm S}^{\rm V}$ as functions of the flux
(measured in units of the flux quantum) and $\beta\omega^{}_{0}$, for a symmetric bridge.
 } \label{SYM}
\end{figure}

When the two leads connected to the electronic reservoirs are identical (making the molecular bridge symmetric) the
 transmission amplitude of the direct bond between the two leads, $t^{}_{o}$ [see Eq. (\ref{TO})], is an even function of $\omega $. The
function ${\cal T}^{\rm V}(\omega ,\Phi )$ given by  Eqs.
(\ref{cf}) and (\ref{TINEL}), is not entirely even or odd in
$\omega $, and therefore {\it a priori} the integrals which give
${\rm X}^{\rm V}_{}(\Phi )$ and $\widetilde{\rm X}^{\rm V}_{}(\Phi
)$ do not vanish. However, the asymmetry in the
$\omega-$dependence of the integrand (which results from the
$\omega-$dependence of the Green function) is not significant. As
a result, ${\rm S}^{\rm V}$ is extremely small, while
$\widetilde{\rm S}^{\rm V}$ is not (because of the extra $\omega$
factor in the integrand), see Fig. \ref{SYM}. These plots are
computed using $\Gamma(\omega )=\Gamma^0\sqrt{1-(\omega /W)^{2}}$,
and $\lambda (\omega )=\lambda^0 [1-(\omega /W)^{2}]$, where $W$
is half the bandwidth, and all energies are measured in units of
$\beta =1/(k^{}_{\rm B}T)$ (we have set $\Gamma^0=\lambda^0 =1$
and $W=50$).

\begin{figure}[ hbtp]
\includegraphics[width=6.cm]{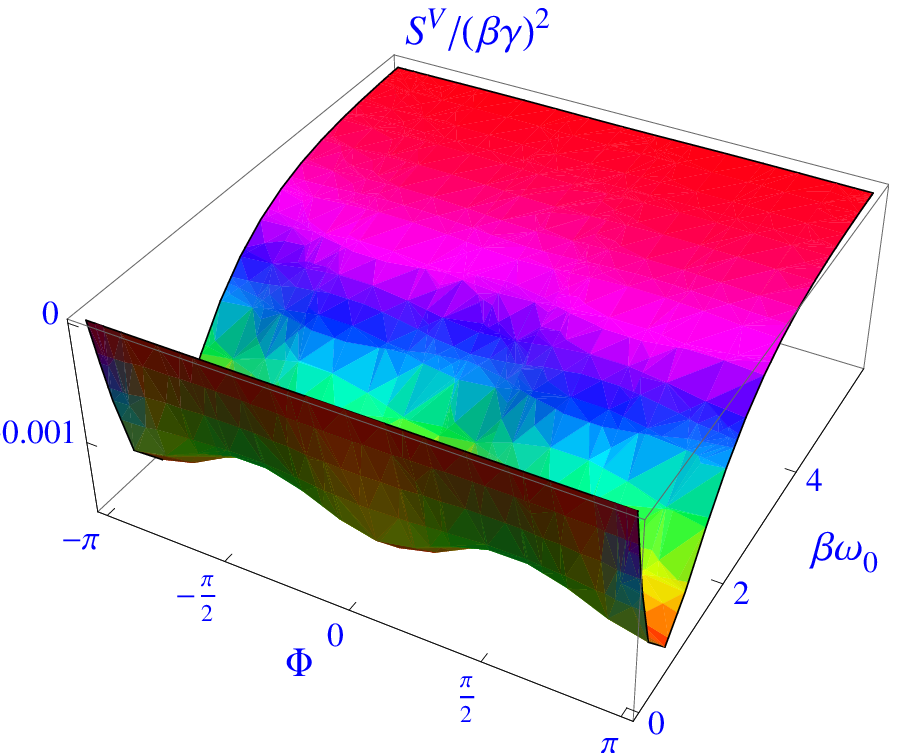}
\includegraphics[width=6cm]{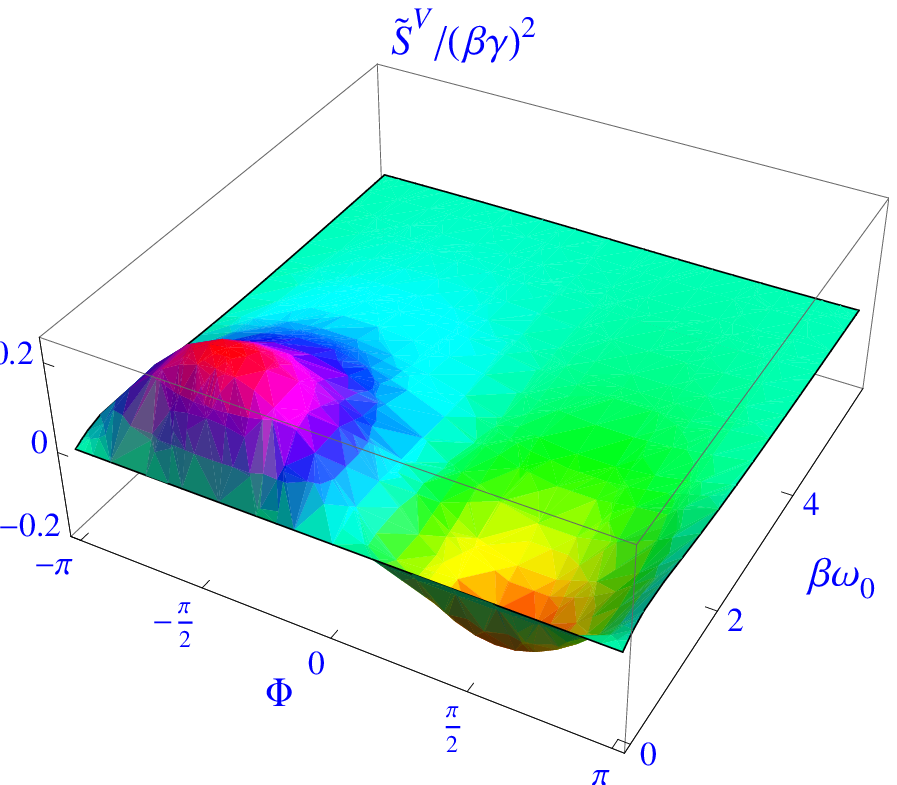}\caption{(color online.) Same as Fig. \ref{SYM},
 for an asymmetric bridge.
 } \label{NONSYM}
\end{figure}

The relative magnitudes of ${\rm S}^{\rm V}$ and
$\widetilde{S}^{\rm V}$ are significantly changed when the
molecule is coupled {\em asymmetrically} to the leads, and
moreover, the character of the  charge carriers on the two
reservoirs is different. Let us  assume that the left reservoir is
represented by an electron band, such that the partial width it
causes to the resonant level is given by
\begin{align}
\Gamma^{}_{\rm L}(\omega )=\Gamma^{0}_{\rm L}\sqrt{\frac{\omega
-\omega^{}_{c}}{\omega^{}_{v}-\omega^{}_{c}}}\ ,\label{GAMLW}
\end{align}
while the right reservoir is modeled by a hole band, with
\begin{align}
\Gamma^{}_{\rm R}(\omega )=\Gamma^{0}_{\rm
R}\sqrt{\frac{\omega_{v}
-\omega^{}_{}}{\omega^{}_{v}-\omega^{}_{c}}}\ .\label{GAMRW}
\end{align}
The corresponding quantity pertaining to the lower arm of the ring in Fig. \ref{sys} is
\begin{align}
\lambda (\omega )=\lambda^0\sqrt{\frac{\omega_{v}
-\omega^{}_{}}{\omega^{}_{v}-\omega^{}_{c}}}\sqrt{\frac{\omega
-\omega^{}_{c}}{\omega^{}_{v}-\omega^{}_{c}}}\ .\label{GAMDW}
\end{align}
Here, $\omega^{}_{c}$ is the bottom of the conductance band (on
the left side of the junction), while $\omega^{}_{v}$ is the top
of the hole band (on the right one).  The energy integration
determining the various transport coefficients is therefore
limited to the region $\omega^{}_{c}\leq\omega\leq\omega^{}_{v}$.
(For convenience, we normalize the $\Gamma$'s by the full band
width, $\omega^{}_{v}-\omega^{}_{c}$.) Exemplifying  results in
such a case are shown in Fig. \ref{NONSYM}, computed with
$\Gamma^0_{\rm L}=\Gamma^0_{\rm R}=\lambda^0 =1$ [see Eqs.
(\ref{GAMLW}), (\ref{GAMRW}), and (\ref{GAMDW})], and
$\omega^{}_{c}=-\omega^{}_{v}=100$, all in units of $\beta$.

Remarkably enough, the coefficient $\widetilde{\rm S}^{\rm V}$,
which measures the ability of the device to turn a temperature
difference on the phonon bath thermalizing the molecular
vibrations into a temperature difference across the molecule is
not very sensitive to the details of the model, apparently because
the term proportional to $\sin\Phi$ in Eq. (\ref{XPT}) is the
dominant one. This means that by reversing the direction of the
magnetic field, one reverses the sign of the temperature
difference or alternatively, the sign of the electronic heat
current. On the other hand, the coefficient ${\rm S}^{\rm V}$,
which sets the scale of the capability to turn a temperature
difference across that phonon bath into an electric current, is
far more sensitive to the details of the model (as expressed e.g.
in our choice for the density of states on the leads) and is less
affected by the magnetic field.

\begin{figure}[ hbtp]
\includegraphics[width=6cm]{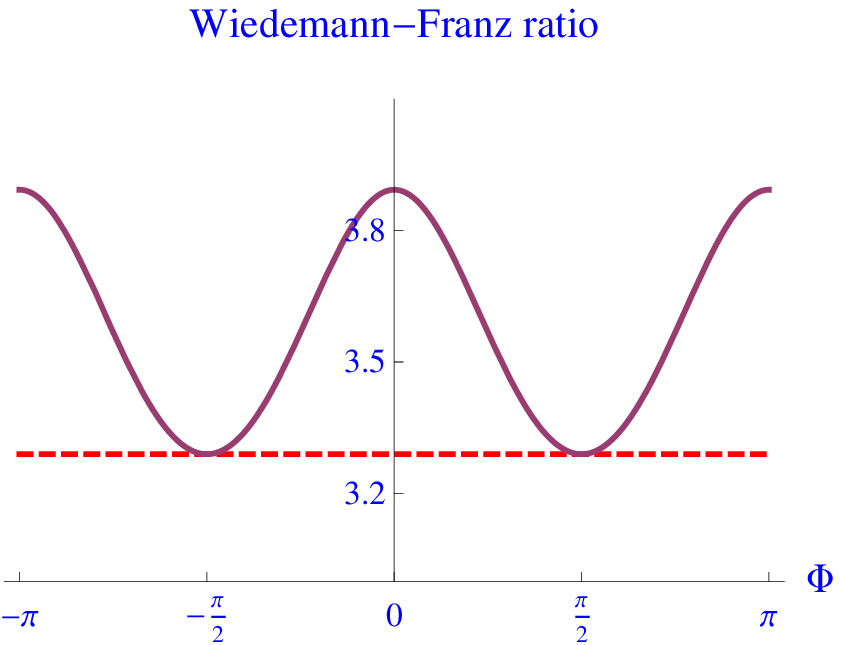}\\
\includegraphics[width=6cm]{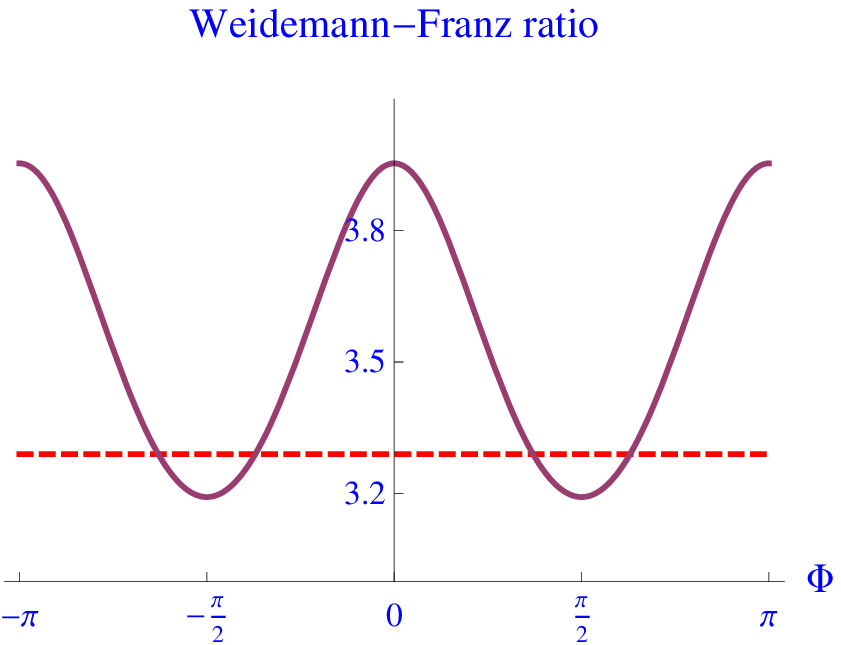}
\caption{ (color online) The Wiedemann-Franz ratio,
$\beta^{2}e^{2}{\rm K}_{2}/{\rm G}$, as a  function of the
magnetic flux (in units of the flux quantum), for a symmetric
(top) and the asymmetric (bottom) bridges. The dotted line is
$\pi^{2}/3$. Parameters are the same as for Figs. \ref{SYM} and
\ref{NONSYM}.
 } \label{WF}
\end{figure}

We now return to the diagonal charge and heat conductances. Figure
\ref{WF} shows the flux-dependence of their dimensionless ratio,
$\beta^{2}e^{2}{\rm K}_{2}/{\rm G}$, plotted with the same
parameters as above. It is interesting to observe that, for both
the symmetric and the asymmetric bridges, this ratio remains quite
close to the textbook Wiedemann-Franz ratio $\pi^2/3$, calculated
from the Fermi-Dirac distribution of a free electron
gas.\cite{Kittel} This is quite surprising, in view of the much
richer resonance structure of the transmission through the ring.

In conclusion, we have found that the themoelectric transport
coefficients through a vibrating molecular junction, placed on an
Aharonov-Bohm interferometer, have an interesting dependence on
the magnetic flux. In particular, the coefficients which relate
the temperature difference between the phonon and electron
reservoirs to the charge and heat currents carried by the
electrons, which exist only due to the electron-vibron coupling,
can be enhanced by the magnetic flux.

\begin{acknowledgments}
Numerous enlightening discussions with Y. Imry are gratefully
acknowledged. This work was supported by the German Federal
Ministry of Education and Research (BMBF) within the framework of
the German-Israeli project cooperation (DIP),and by  the US-Israel
Binational Science Foundation (BSF).
\end{acknowledgments}


\begin{thebibliography}{999}
\bibitem{COM1}
The use of this identity is justified by the assumption that the reservoirs supplying the particles are kept at thermal equilibrium.
\end{thebibliography}
\end{document}